\documentclass[sigconf, screen]{acmart}

\usepackage{xcolor}
\usepackage{rotating}
\usepackage{booktabs}
\usepackage{multirow}
\usepackage{array}

\setcopyright{none}

\begin{document}
\title[Stitching Well-being]{Stitching Well-being: Negotiating Embodied Engagement with an Embroidery Toolkit among Female STEM Students}

\author{Xinglin Sun}
\orcid{0009-0003-8752-202X}
\affiliation{
\department{College of Design and Innovation}
  \institution{Tongji University}
  \city{Shanghai}
  \country{China}}
\email{xinglinsun@tongji.edu.cn}

\author{Yifan Yan}
\affiliation{
\department{College of Design and Innovation}
  \institution{Tongji University}
  \city{Shanghai}
  \country{China}}
\email{yifanyan@tongji.edu.cn}

\author{Yifan B. Feng}
\orcid{0009-0009-9589-7589}
\affiliation{%
\department{Dyson School of Design Engineering}
  \institution{Imperial College London}
  \city{London}
  \country{United Kingdom}}
\email{yifanfeng4@acm.org}

\author{Shengsi Li}
\affiliation{
\department{College of Design and Innovation}
  \institution{Tongji University}
  \city{Shanghai}
  \country{China}}
\email{2633449@tongji.edu.cn}

\author{Caroline Claisse}
\orcid{0000-0003-4002-6136}
\affiliation{
\department{Open Lab}
  \institution{Newcastle University}
  \city{Newcastle upon Tyne}
  \country{United Kingdom}}
\email{caroline.claisse@newcastle.ac.uk}

\author{Qi Wang}
\authornote{Corresponding author}
\orcid{0000-0002-2688-8306}
\affiliation{
\department{College of Design and Innovation}
  \institution{Tongji University}
  \city{Shanghai}
  \country{China}}
\email{qiwangdesign@tongji.edu.cn}

\author{}
\orcid{}
\affiliation{%
\department{}
\institution{} 
\city{} 
\country{}}
\email{}

\author{}
\orcid{}
\affiliation{%
\department{}
\institution{} 
\city{} 
\country{}}
\email{}

\renewcommand{\shortauthors}{Sun et al.}

\begin{abstract}
% Embodied interaction and soma-oriented design foreground bodily experience as a source of knowledge and reflection, yet how people develop and sustain embodied engagement through designed practices remains less understood. We present an exploratory design study of an Embodied Embroidery Toolkit with 11 female STEM students. Across two five-day toolkit-use phases, we combined interviews, journals, participant-created artifacts, and individual physiological measures to examine how participants engaged with and appropriated embroidery, sensory materials, and optional e-textiles. Our findings show that participants developed diverse and personally meaningful ways of engaging with embodied embroidery. They negotiated structure, used repetitive stitching to support attention and reflection, materialized bodily experiences through somatic anchors, and adapted materials and technologies to their bodies, identities, disciplinary orientations, and everyday contexts. Bringing data sources into dialogue revealed converging physiological, self-reported, and lived accounts alongside different temporal dimensions of engagement. We contribute an account of embodied engagement as situated negotiation and articulate design sensibilities for scaffolding rather than prescribing engagement, supporting diverse attentional possibilities, making bodily experiences tangible and revisitable, and enabling appropriation.

Embodied interaction and soma-oriented design foreground bodily experience as a source of knowledge and reflection, yet tend to treat engagement as something a design elicits, leaving open how people themselves develop and sustain it. We present an exploratory design study of an Embodied Embroidery Toolkit with 11 female STEM students across two five-day use phases, combining interviews, journals, artifact making, and physiological measures. Participants developed diverse and personally meaningful ways of engaging: negotiating the structure the toolkit offered, using repetitive stitching to steady attention and open space for reflection, materializing bodily experience through somatic anchors, and adapting materials and optional e-textiles to their bodies, identities, and everyday routines. We contribute an account of embodied engagement as situated negotiation, in which people continually configure relationships among attention, bodily experience, materials, and context, and articulate design sensibilities for scaffolding rather than prescribing engagement, supporting diverse attentional possibilities, and enabling appropriation. 
\end{abstract}

%https://dl.acm.org/ccs 

%\ccsdesc[300]{Social and professional topics~Women}


\begin{teaserfigure}
    \centering
    \includegraphics[width=\linewidth]{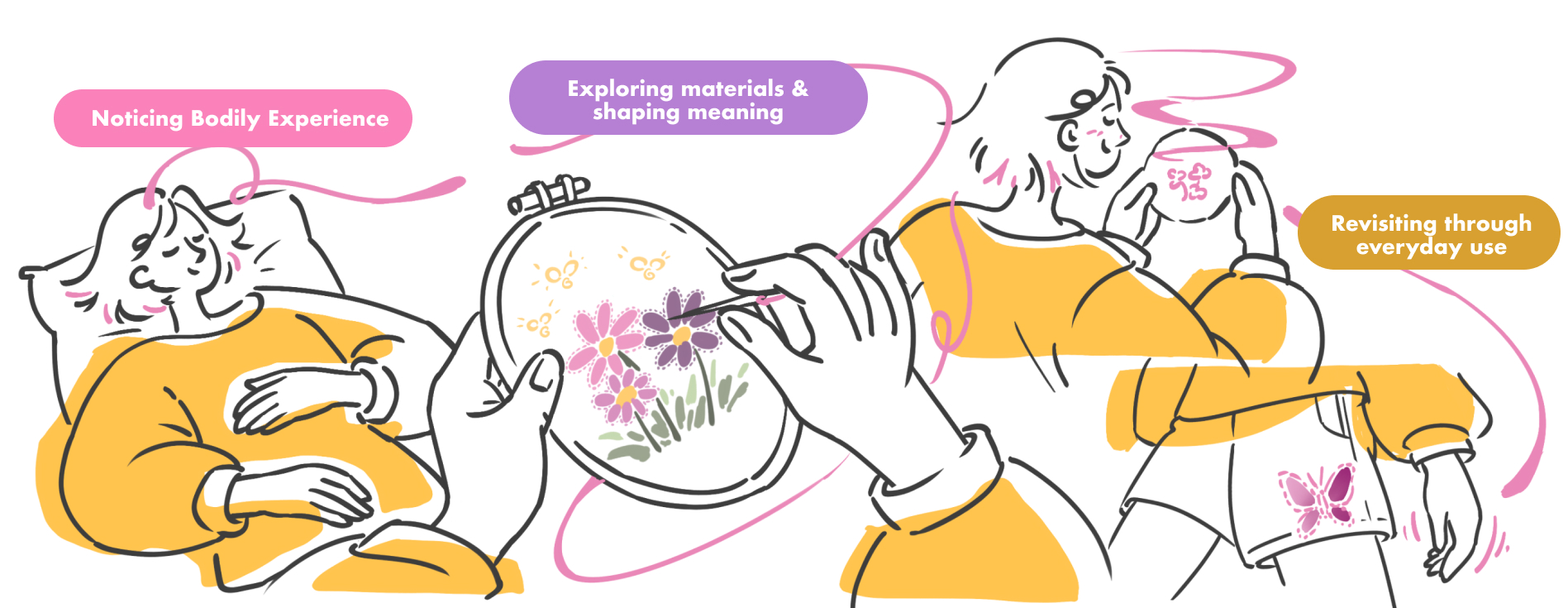}
   \caption{Somatic anchoring as negotiated embodied engagement across time and context. The illustration synthesizes how participants noticed bodily experiences, explored embroidery and sensory materials such as color-changing, luminous, and scented threads, and shaped personally meaningful sensory and visual forms that could be touched, smelled, viewed, worn, or carried into everyday life, connecting moments of reflection and making with subsequent encounters with the body.}
    \Description{An illustration connecting bodily noticing, embroidery making, and everyday use. On the left, a person reclines with closed eyes and a hand resting on their abdomen, representing attention to bodily sensations. In the center, hands stitch an illustration of small flowers, blades of grass and fireflys in an embroidery hoop. On the right, a person holds an embroidered object near their nose, while a lower view shows embroidery on clothing beside a hand. Flowing pink lines connect the scenes. Together, the scenes illustrate how bodily experiences are explored through making and given material forms that can be encountered again through sensory engagement in everyday life.}
\end{teaserfigure}

\maketitle

\section{Introduction}
Women in STEM education experience a range of challenges to their mental well-being, including stress, anxiety, and depression, shaped not only by academic demands but also by disciplinary cultures, gendered expectations, and students' perceptions of themselves \citep{settles2016climate, beroiza2025gendered, gao2020gender, reidy2025mental}. Although large-scale studies have established important relationships between gender, disciplinary context, and mental well-being, they provide limited insight into how these conditions are lived and experienced in everyday educational settings. In particular, less is understood about how female STEM students notice, interpret, and respond to their bodily and emotional experiences as they navigate academic and disciplinary demands.

Embodied interaction and soma-oriented design offer a different way of approaching this question by positioning bodily experience as a source of knowledge, reflection, and design inquiry \citep{dourish2001where, hook2018soma}. Rather than treating the body primarily as an object to be measured or acted upon, embodied approaches attend to how people perceive, make sense of, and transform their experiences through bodily engagement. Such perspectives have increasingly informed HCI research on women's health, including reproductive, sexual, and mental well-being \citep{park2025designing, hua2023embodied, sun2025conversations}. However, embodied and somatic approaches remain less explored as everyday practices for women's mental well-being, particularly within technical and computational educational contexts. This raises a question to designers: \textit{how might we support embodied engagement without prescribing what participants should feel, notice, or do?}

Textile craft offers a particularly promising context for exploring this question. Embroidery involves sustained hand movement, tactile materials, repetition, and close attention to the formation of stitches, while craft practices have also been historically and culturally associated with women's everyday lives \citep{classen_deepest_2012}. Recent HCI research has explored textile making as a space for agency, reflection, self-expression, and feminist practice \citep{ermer2023craftivism, 10.1145/3715336.3735412}. The emergence of expressive e-textiles further creates opportunities to connect bodily experience with responsive materials and interactive technologies \citep{bolesnikov2023wearable, walczak2025expresense, lyu2026reimagining}. Yet how such a practice is actually taken up by women in technical educational contexts, and how they negotiate its structure, materials, and technological possibilities in relation to their own bodies and everyday lives, remains underexplored.

To investigate this space, we present an exploratory design study of an \textbf{Embodied Embroidery Toolkit} with 11 female STEM students. The toolkit combined conventional embroidery materials with body-mapping and reflective activities, sensory materials, and optional e-textile components. Rather than treating the toolkit as a fixed therapeutic intervention, we used it as a design probe \citep{wallace_making_2013} through which participants could explore different ways of attending to, expressing, and engaging with bodily and emotional experiences. Across two five-day toolkit-use phases, we collected interviews, reflective journals, participant-created artifacts, self-report measures, and individual physiological traces, examining both cross-case patterns and the unfolding of particular embodied experiences.

Our findings show that embodied engagement developed through participants' ongoing negotiation with the practice. Participants negotiated different degrees of structure, used repetitive stitching to organize attention in different ways, materialized bodily experiences through personally meaningful somatic anchors, and appropriated materials and technologies according to their bodies, identities, disciplinary orientations, and everyday contexts. Drawing these data sources into dialogue further revealed how physiological traces, self-reports, and lived accounts could converge around particular experiences while also capturing different temporal dimensions of engagement.

Beyond reporting these empirical insights, we use them to articulate \textbf{design sensibilities} as transferable, intermediate-level design knowledge rather than prescriptive rules \citep{gamboa2026sensibilities}. Design sensibilities offer a way of carrying situated insights from particular design encounters into other contexts while retaining their relational and situated character. In our case, they help translate participants' diverse experiences into considerations for designing embodied interactions without reducing those experiences to a single desired outcome.

This paper makes two contributions. First, we provide an empirical account of \textbf{embodied engagement as situated negotiation}, showing how participants actively configure relationships among attention, bodily experience, materials, technologies, and context. Second, we articulate \textbf{four design sensibilities for supporting such negotiation: scaffolding without prescribing, supporting multiple ways of attending, making embodied experience tangible and revisitable, and leaving room for situated appropriation}. Together, these contributions extend embodied interaction research by shifting attention from designing for a particular embodied outcome toward designing conditions in which people can develop their own ways of engaging.

\section{Related Work}
%In this section, we first outline the mental and psychological well-being challenges faced by students in STEM (science, technology, engineering, maths), arguing that the distinct pressures for females in these fields make their well-being a critical question for the HCI and design community. We then review scholarship on embodied practice, reframe gendered bodily experience as a resource for understanding mental well-being, and discuss the role of body-centric, somatic methods in exploring the design space to support female STEM students. Finally, we examine creative health research on crafting and textile practice and highlight the opportunity to use embodied embroidery practice to bridge gender, bodily practice, and mental health work.

\subsection{Mental Well-being Among Women in STEM}
Undergraduate students in STEM fields experience substantial mental health challenges, including anxiety and depression, which have been associated with cultures of intense competition and academic achievement \citep{lipson2016major, posselt2016competition, asghar2024mental}. For women in STEM, in particular, \citet{settles2016climate} highlight that a negative academic climate, greater interference between women's identities as women and as scientists, and lower perceptions of their own scientific performance can all potentially result in poorer psychological and mental well-being. 

Recent large-scale studies have therefore begun to examine how gender and disciplinary context jointly shape STEM students' mental well-being, calling for more gender-sensitive approaches to well-being support. Given the well-documented gender disparities in STEM education \citep{eddy2016beneath}, particularly the discussion on women's lower sense of belonging and academic self-efficacy \citep{sanchez2021comparing}, as well as gendered barriers to mental health support more broadly \citep{RCPsych2026WomenMentalHealth}, several recent studies have examined whether and how gender shapes undergraduate STEM students' mental health at large scale, and has called for more gender-sensitive interventions to ensure equity in well-being support. 

Women and non-binary STEM students have been found to report higher levels of depression and stress than men, alongside lower academic self-efficacy, pointing to self-efficacy as an important resource for coping with psychological distress \citep{beroiza2025gendered}. Similar gendered patterns have been reported among Chinese STEM students, where women show particularly high levels of anxiety, with body image, drinking habits, and academic performance further shaping their experiences \citep{gao2020gender}. Analysis of nationwide US survey data likewise suggests that both gender and STEM disciplinary status are associated with anxiety and depression \citep{pigart2025stem}. Importantly, these patterns are not consistent across STEM contexts: \citet{reidy2025mental}, for example, found that women in gender-balanced STEM majors reported more anxiety, depression, and trauma symptoms than those in male-dominated majors. Together, these findings suggest that women's well-being in STEM cannot be understood through gender alone, but emerges through its intersection with disciplinary environments, academic expectations, and students' perceptions of themselves.

Overall, these studies establish important relationships among gender, disciplinary context, and STEM students' mental well-being. However, large-scale quantitative research provides limited insight into how these conditions are lived and experienced in everyday STEM settings. In particular, we know less about how gendered expectations, disciplinary practices, and academic pressures become entangled with students' everyday bodily and emotional experiences. Rode and Poole, for example, show that engagement with digital technologies is intertwined with gender identity and technical identity, challenging accounts that explain technology engagement primarily through technical expertise \citep{10.1145/3196839.3196845}. Building on this perspective, recent work calls for approaches that are attentive to the diversity of women's experiences in STEM \citep{rode2025reframing}.This points to the need for more situated approaches that examine how female STEM students experience and navigate the conditions shaping their well-being.

\subsection{Embodiment and Somatic Perspectives for Women's Well-being Research in HCI}
Early cognitive science largely conceptualized cognition as internal representation and computation, with the body treated primarily as an input-output system \citep{neisser1967cognitive, simon1969sciences}. In contrast, embodied and enactive approaches understand cognition as emerging through ongoing bodily engagement with the environment \citep{varela1991embodied, wilson2002six}. This perspective foregrounds the lived body as constitutive of perception, action, and meaning-making, shifting attention from what people cognitively represent to how they experience and make sense of the world through their bodies.

In HCI, these perspectives have reframed interaction as situated in bodily practice rather than as an exchange of symbolic representations \citep{dourish2001where}. Subsequent work has further examined bodily action as a resource for perception, cognition, and engagement with interactive systems \citep{kirsh2013embodied, feltham2014slow, nakagaki2016materiable, rogers2006framework}. Soma design extends this perspective by drawing on somaesthetics to position first-person bodily experience as a source of knowledge, creativity, and transformation, while offering methods for cultivating such sensibilities in designers and participants \citep{hook2018soma}. Together, these approaches position the body not simply as a site where interaction occurs, but as a medium through which people perceive, attend to, interpret, and transform their experiences.

These embodied and somatic perspectives have increasingly informed HCI and design research on women's health. Existing work has explored bodily experience in areas including reproductive health \citep{park2025designing, sondergaard2021designing, hofer2025my, hess2026sensing, ciolfi2021resisting}, sexual health \citep{teisanu2025designing, hua2023embodied}, and mental well-being \citep{sun2025conversations}. Across these contexts, researchers attend to how women feel, move, perceive, and make meaning through their bodies, challenging approaches that treat the body primarily as an object to be measured, monitored, or technologically acted upon. 

However, embodied and somatic approaches remain underexplored in women's everyday mental well-being, particularly in educational and working contexts. This gap is particularly relevant to female STEM students, whose well-being is shaped by intertwined psychological, academic, disciplinary, and gendered conditions. We therefore explore embodied practice as a way for female STEM students to notice, reflect on, and give material form to their everyday bodily and emotional experiences, turning to textile craft as a tangible and culturally situated form of embodied practice.

\subsection{Textile Crafts as Women's Mental Well-being Support}
HCI researchers have increasingly explored embodied creative practices such as drawing \citep{meng2026creating}, dancing \citep{piitulainen2022vibing, tokida2026immersivedmt}, knitting \citep{jiang2026loopscroll, pschetz2013movement}, knotting \citep{tsaknaki2024knotting}, and weaving \citep{recupero2021enabling} as resources for supporting women's mental well-being. Within this broader body of work, textile crafts are of particular interest to our research because of their long-standing relationship with women's everyday lives and practices.

Across cultures, making by hand and craftwork have long been gendered as ``women's work'' \citep{classen_deepest_2012}. Although this framing has historically devalued craft and reinforced gendered divisions of labor, its intergenerational presence among women has also contributed to its familiarity and cultural accessibility. Feminist craftivism has further challenged the marginalization of craft by reclaiming traditionally feminized making practices as forms of feminist expression, critique, and collective action \citep{ermer2023craftivism}. Related work on feminist DIY culture has similarly explored how making can provide women with ways to construct intimate technologies and negotiate relationships between their bodies, materials, and everyday lives \citep{10.1145/3715336.3735412}. Rather than treating craft as merely domestic or decorative practice, this body of work positions making as a site for agency, knowledge production, and self-expression.

Textile making can also support personal reflection, creativity, and emotional expression. Material engagement provides opportunities to externalize, revisit, and give form to experiences that may otherwise remain difficult to articulate. \citet{10.1145/3803784.3809284}, for example, uses autoethnography alongside an emerging textile art practice to examine how creative work is shaped through material, cultural, and embodied contexts. Her account positions textile making as a situated and relational practice in which reflection can make otherwise subtle aspects of creative experience visible. In HCI, this perspective resonates with work that understands making as a means through which women can develop agency, reflect on personal experiences, and express themselves creatively \citep{ermer2023craftivism}.

This interest in expressive making has increasingly extended to expressive wearables and e-textiles. Expressive wearables afford creative self-reflection and self-expression by allowing people to modify, reinterpret, and communicate bodily experiences through interactive features and tangible forms. HCI research has explored these possibilities in relation to the gendered relationship between bodies and technology \citep{bolesnikov2023wearable} and to broader personal and cultural identities in everyday practice \citep{epp2020identity, walczak2025expresense}. Recent work on musical performance designed an expressive e-textile interface that challenged the masculine-coded emphasis on precision, mastery, and control in digital musical instruments \citep{lyu2026reimagining}. It shows how redistributing agency between performer and instrument can foster more reflective engagement with bodily movement and expression.

This literature positions textile making as more than a means of producing an artifact. It can provide a material and embodied space for agency, creativity, reflection, and self-expression, while expressive e-textiles further open possibilities for connecting bodily experience with interactive materials. However, existing work has paid less attention to how such practices are taken up by women in technical and computational educational contexts, where making may encounter different habits of precision, problem-solving, and technological engagement. Our work builds on textile craft and expressive e-textile research to explore how an embodied embroidery practice can become a personally meaningful space for reflection, bodily engagement, and creative expression among female STEM students.

\section{Designing Embodied Embroidery Toolkit}

\begin{figure}[!ht]
    \centering
    \includegraphics[width=\linewidth]{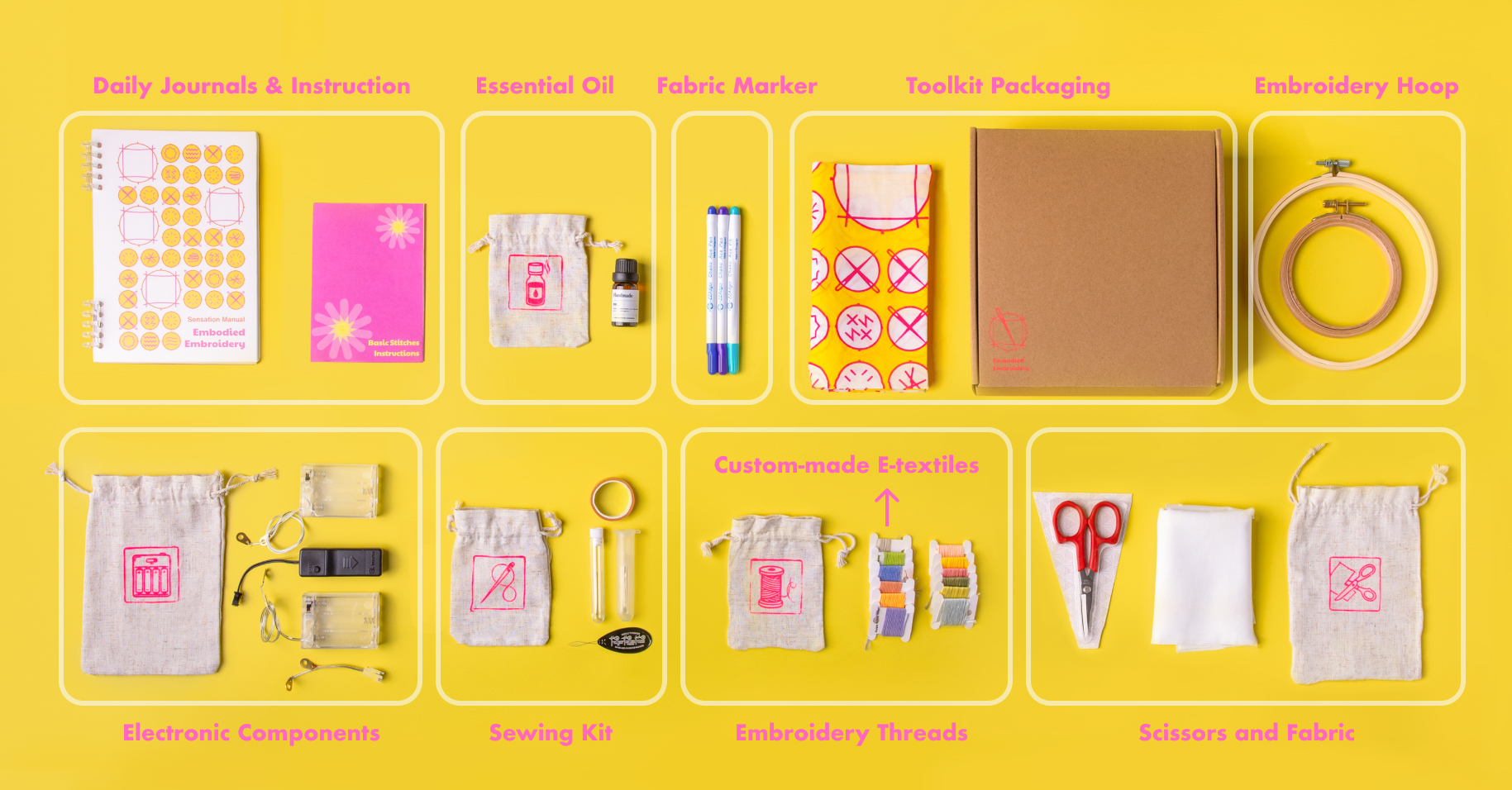}
 \caption{Overview of the Embodied Embroidery Toolkit, highlighting its customised interactive threads alongside conventional and sensory embroidery materials}
    \Description{Flat-lay photograph of an embodied embroidery toolkit arranged in labeled sections on a bright yellow background. The toolkit includes daily journals and instructions, essential oil, three water-erasable pens, patterned fabric and cardboard packaging, two embroidery hoops, electronic components and battery packs, a sewing kit, multicolored embroidery and e-textile threads, red-handled scissors, white fabric, and several labeled drawstring pouches.}
    \label{fig:tookit}
\end{figure}

\subsection{Toolkit Rationale}
We conducted an exploratory study using an embodied embroidery toolkit for exploring well-being among female STEM students. Rather than seeking a fixed intervention, we treated the toolkit as a design probe \citep{wallace_making_2013} for exploring how participants engage with embodied textile practices through which embodied, material, and sensory possibilities could be explored. The development of the toolkit builds on \citet{sun2025conversations}'s research on embodied approaches to stress self-disclosure among adolescent girls and their subsequent exploration of embroidery as an embodied and reflective practice for women’s mental well-being \citep{sun2026reclaiming}. Together, these studies suggested opportunities for using embroidery not only as a craft activity, but also as a material medium through which bodily experiences could be noticed, expressed, and revisited.

\subsection{Design Objectives}
Our toolkit was intended to provide a structured entry point into embodied making while leaving room for participants to determine what to attend to, represent, and make. Rather than prescribing a particular relationship between a material and a psychological state, we sought to create conditions in which bodily sensations and stress-related experiences could be explored through stitching, materials, and optional interactive technologies. Our design objectives include: 

\begin{itemize}
    \item \textbf{Support embodied attention.} We aimed to make bodily engagement an integral part of the activity. Unlike screen-based hand-eye coordination, embroidery requires sustained hand movements and close engagement with fabric, thread, and stitch formation, offering opportunities for attention to be organized through repetitive and tactile making.
    \item \textbf{Enable personal expression and materialization.} Building on the previous research on body mapping and embodied metaphors \citep{anne_cochrane_body_2022, dauden_roquet_body_2020,sun2025conversations}, we aimed to encourage participants to translate bodily sensations and experiences into personally meaningful visual and material forms. The toolkit was designed to support participants in developing their own relationships between experience, motif, color, texture, and stitch.
    \item \textbf{Open multi-sensory pathways.} Conventional embroidery is primarily tactile and visual. We extended the material repertoire to include sensory materials that could introduce qualities associated with temperature, smell, and light. These materials were intended to open alternative ways of engaging with bodily experience rather than to function as predefined mechanisms for self-regulation.
    \item \textbf{Support situated appropriation.} Finally, we wanted embodied making to extend beyond the immediate research setting. The toolkit therefore included materials that could be incorporated into portable, wearable, or everyday objects. This enabled participants to determine how and where embodied textile interactions might become meaningful within their daily lives.
\end{itemize}

Overall these objectives positioned the toolkit as a repertoire of possibilities rather than a fixed intervention. The design provided enough structure to make embodied embroidery approachable while retaining openness regarding the form, meaning, sensory qualities, and context of use.

\subsection{Material Choices}
We translated these objectives into a combination of conventional embroidery materials, printed materials, sensory materials, and optional electronic components (see Table~\ref{tab:design-objectives}). The selection was intended to create different possibilities for engaging with the body while maintaining embroidery as the central making practice. We also customized two types of interactive threads for hand embroidery. First, as commercially available thermochromic threads are primarily designed for machine embroidery and are relatively fine for hand stitching, we braided multiple strands of thermochromic embroidery thread around silver conductive thread with good heating performance. This increased the thread's thickness while allowing it to respond to heat during use. Second, we braided loosely structured, multi-strand cotton thread with relatively large gaps around the same type of silver conductive thread. The resulting braided thread was soaked as a whole in an essential oil of the participant's choice and left to dry before use. When electrically heated through the conductive core, the cotton fibers gradually released the absorbed fragrance.

\begin{table}[!htbp]
\centering
\caption{Design Objectives and Material Choices}
\label{tab:design-objectives}

\renewcommand{\arraystretch}{1.8}
\begin{tabular}{p{0.21\linewidth} p{0.29\linewidth} p{0.42\linewidth}}
\toprule
\textbf{Design objective} & \textbf{Material choices} & \textbf{Design rationale} \\
\midrule

\textbf{Support embodied attention}
& Embroidery hoop, sewing kit, embroidery threads, basic stitches introduction zine
& Provide portable, hands-on engagement with embodied-making
\\

\textbf{Enable personal expression}
& Body maps, daily journals, body-map inspired motifs
& Allow participants to translate bodily sensations and experiences into personally meaningful forms
\\

\textbf{Support multi-sensory engagement}
& Tactile, scented, thermochromic, and luminous threads
& Open different sensory pathways through touch, smell, temperature, and visual change
\\

\textbf{Support situated use}
& Wearable and portable forms, e.g., clothing, bags, and accessories
& Enable participants to carry and incorporate their creations into everyday contexts
\\

\textbf{Offer optional interactive possibilities}
& Lightweight electronic components with garment clips 
& Allow participants to explore responsive textile interactions without making technology a prerequisite
\\

\bottomrule
\end{tabular}
\end{table}

The final material collection combined textile, sensory, and interactive possibilities, allowing embodied engagement to emerge through different configurations of materials rather than through predefined interaction.

\subsection{Design Iterations}
The toolkit was developed through iterative design, drawing on our team members' expert knowledge, with particular attention to how different forms of technology might shape participants’ engagement with embroidery and embodied reflection.

\textbf{Moving away from screen-based programming.} In an early iteration, we considered inviting participants to use basic programming as part of the toolkit. This direction was motivated by the possibility of connecting embroidery with computational making and allowing participants to create their own interactive textile artifacts. However, we did not take this approach because programming would require participants to return to screens and shift their attention toward screen-based tasks. This conflicted with our initial intention of using embroidery as a slow, hands-on practice that could provide an alternative to screen-based routines. We therefore removed programming as a required part of the toolkit and retained embroidery as the primary mode of making.

\textbf{Separating e-textile exploration from initial embodied practice.} We also initially planned to introduce e-textile materials during Phase 1, alongside the basic embroidery activities. Early consideration of this configuration suggested that conductive materials, wiring, and electronic components could introduce additional technical and material problems for participants to solve. Rather than attending to bodily sensations, embroidery, or reflection, participants could become primarily occupied with making the technology work. We therefore moved e-textile exploration to the later prototyping stage. Phase 1 focused on embroidery, body scanning, body mapping, and reflective activities, while e-textile and sensory materials were introduced as optional resources when participants developed their own prototypes in Workshop 2.

These iterations resulted in a toolkit in which embroidery formed the core practice, while technology remained an optional extension rather than a prerequisite for participation. The final toolkit combined embroidery materials, reflective resources, and optional sensory and e-textile components (see Figure~\ref{fig:tookit}).

\section{Methods} \label{methods}
% purposive sampling, poster, visit, online and in-person 
% participant table 

\subsection{Participants} \label{participants}
We recruited female STEM students through purposive sampling, seeking participants whose educational backgrounds were relevant to our exploration of embodied textile practice in technology-oriented educational contexts. Recruitment was conducted through multiple channels, including recruitment posters, in-person visits to relevant student settings, and online recruitment through emails and notice boards. Recruitment materials introduced the study as an exploration of embroidery, embodied experience, and well-being, and provided information about the expected activities and time commitment.

Eleven participants aged 19-23 years completed the study ($M$ = 21.27, $SD$ = 1.85) (see Table~\ref{tab:participants}). Participants came from a range of computing-related disciplines, including Data Science, Cyber Security, Computer Science, and Computing and Mathematics. Their prior experience with embroidery and textile crafts varied from no experience to regular craft practice, while their familiarity with reflective practices also varied. This variation provided a range of familiarity with both the material practice and reflective activities introduced in the study.

All participants self-identified as female except for P4, who described their gender identity as ``in between women and non-binary''. We retain this participant's self-description rather than imposing a categorical label. This was particularly important to our analysis, which did not assume a singular or uniform experience of gendered embodiment.

Participants who met the criteria and expressed interest were provided with further information about the study before enrollment. No participant was selected based on a particular level of stress or expected response to the toolkit. Ethical procedures and informed consent are described in Section 4.2.

\begin{table}[!htbp]
\centering
\caption{Participant characteristics.}
\label{tab:participants}
\resizebox{\columnwidth}{!}{%
\begin{tabular}{lclll}
\toprule
\textbf{ID} & \textbf{Age} & \textbf{Field of Study} &
\textbf{Embroidery/Textile Experience} &
\textbf{Prior Reflective Practice} \\
\midrule
P1 & 23 & Data Science & None & None \\
P2 & 23 & Data Science & Some experience & Occasional \\
P3 & 23 & Data Science & Beginner & None \\
P4 & 23 & Cyber Security & Beginner & Occasional \\
P5 & 20 & Computer Science & Some experience (crochet) & None \\
P6 & 19 & Computer Science & None & Occasional \\
P7 & 19 & Computer Science & None & Occasional \\
P8 & 22 & Computer Science& Beginner & Occasional \\
P9 & 23 & Data Science & None & Occasional \\
P10 & 20 & Computer Science & Beginner & Occasional \\
P11 & 19 & Computing and Mathematics & Regular practice (crochet) & None \\
\bottomrule
\end{tabular}%
}
\end{table}

\subsection{Ethics and Additional Considerations} \label{ethics}
This study received ethical approval from the lead/first author's institution and was cross-registered at the collaborative authors' affiliations. Each participant was compensated with a \$40 voucher and encouraged to keep the toolkit as a token of their participation. In addition to following relevant institutional requirements and the ACM ethical guidelines, we incorporated additional ethical considerations. 

% \textbf{Recruitment, selection, and consent.} Recruitment materials provided prospective participants with information about the research purpose, expected activities, time commitment, and data collection procedures. Before participation, all participants received an information sheet describing the study and were given an opportunity to ask questions. Written informed consent was obtained from all participants before data collection. Participants were informed that participation was voluntary and that they could withdraw from the study at any point, without providing a reason. Participants who finish the study will receive a voucher for their contribution.

\textbf{Treatment during the study.} Because the study involved reflection on stress and bodily experience, we took care not to frame the toolkit as a clinical or therapeutic intervention. Participants were not required to disclose personal experiences of stress or other sensitive information. During the study, participants could choose whether and how to engage with particular activities and materials, and could pause, skip, or discontinue an activity when needed. Researchers checked in with the participants in person once in the study, and remained available throughout the study to answer questions and respond to participants' concerns and needs.

The study also recognized that embodied practices do not necessarily produce the same experience for everyone. We therefore avoided treating calmness, bodily awareness, or stress reduction as expected outcomes of participation. Participants' divergent, uncomfortable, or non-beneficial experiences were treated as valid accounts of engagement rather than as failures to follow the activity.

% \textbf{Data handling, privacy, and confidentiality.} All research data were pseudo-anonymized during analysis and reporting. Participants were assigned identifiers (e.g., P1, P2, and P4), which are used throughout this paper instead of names. Identifying information was removed and was not included in the analysis dataset used for reporting. Research data were stored in secured university server and were accessible only to authorized research team members. 

\subsection{Study Design} \label{study_design}
\begin{figure}[!ht]
    \centering
    \includegraphics[width=\linewidth]{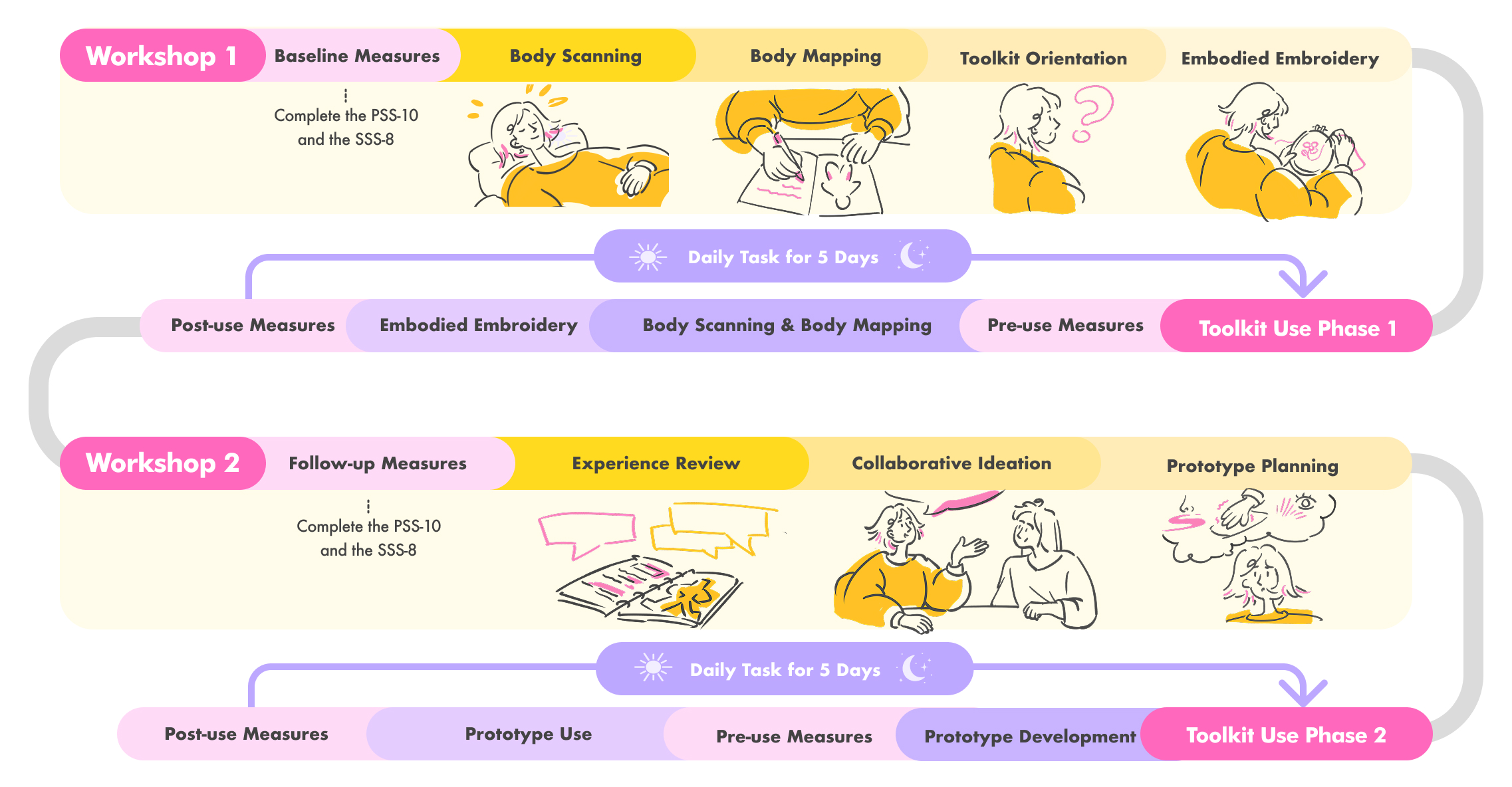}
    \caption{Longitudinal and iterative study design combining repeated embodied engagement, multimodal measurement, and participant-led prototyping.}
    \Description{Flowchart of the study procedure. Workshop 1 consisted of baseline PSS-10 and SSS-8 measures, body scanning, body mapping, toolkit orientation, and embodied embroidery. During Toolkit Use Phase 1, participants completed five days of daily pre-use measures, body scanning and mapping, embodied embroidery, and post-use measures. Workshop 2 included follow-up PSS-10 and SSS-8 measures, an experience review, collaborative ideation, and prototype planning. During Toolkit Use Phase 2, participants developed their prototypes and completed five days of pre-use measures, prototype use, and post-use measures.}
    \label{fig:procedure}
\end{figure}

We conducted a two-cycle study combining facilitated exploration of the Embodied Embroidery Toolkit with subsequent periods of individual use. The study consisted of two workshops, each followed by a five-day toolkit-use phase, and subsequent interviews. The first cycle focused on participants' experiences of embodied textile making, while the second cycle examined how participants appropriated the toolkit and their own prototypes in everyday contexts (see Figure~\ref{fig:procedure}) .

\paragraph{Workshop 1 Structure.}
Participants took part individually or in small groups of two to three, depending on their preference. Participants first completed the Perceived Stress Scale (PSS-10)~\cite{cohen1983global} and the Somatic Symptom Scale-8 (SSS-8)~\cite{gierk2014somatic} as baseline measures before taking part in the study. Participants then learned basic sewing and embroidery techniques and spent approximately one hour creating their first embodied embroidery piece \cite{sun2026reclaiming} after engaging with body scanning and body mapping activities. Participants were introduced to the reflective notebook and the different embroidery, sensory, and e-textile materials available in the toolkit that would be used in the study.

\paragraph{Toolkit Use Phase 1.}
Following Workshop 1, participants used the toolkit for five days in their everyday settings. They were invited to engage with activities including body scanning, body mapping, embodied embroidery, and reflective journaling at their own pace. Participants also recorded heart rate variability (HRV), heart rate, stress level, and respiratory measurements before body scanning and after embroidery activities.

\paragraph{Workshop 2 Structure.}
Participants met with the researcher individually or in pairs if they prefer and filled in the PSS-10 and SSS-8 for the second time. They then reviewed their experiences from Toolkit Use Phase 1, bringing the embodied embroidery pieces they had created and their reflective notebooks to the session. Participants were invited to revisit and discuss their observations, bodily experiences, and reflections from the previous five days. The researcher and each participant then engaged in a collaborative ideation process to explore how they might extend, transform, or respond to their Phase 1 experiences through an embroidery-based prototype. Together, they discussed possible forms, materials, sensory qualities, and modes of use, with the researcher providing support in translating participants' reflections and ideas into a feasible design for Phase 2.

\paragraph{Toolkit Use Phase 2.}
Following Workshop 2, participants developed their own prototypes based on the ideas and directions developed during the session. They then engaged with these prototypes over a further five-day period in their everyday settings. Participants were encouraged to use, adapt, and modify their prototypes as appropriate, while documenting their experiences on the journal. HRV, heart rate, stress level, and respiratory measurements were collected before and after prototype use.

\paragraph{Post hoc Interviews.}
Following the toolkit-use phases, participants took part in two-part interviews. The first part was a semi-structured interview focused on how participants conceptualized, developed, and interpreted their embodied embroidery pieces and prototypes, including their design rationales, material choices, interactions, and reflections on bodily and emotional experiences. The second part adopted a micro-phenomenological interview approach \cite{petitmenginDescribingOnesSubjective2006} to investigate participants' lived experiences during specific moments of engaging with the toolkit. Participants were invited to recall a particular moment that remained noticeable, meaningful, calming, frustrating, or otherwise salient in the study, and were guided to describe its temporal unfolding, bodily sensations, tactile and material engagement, shifts in attention, and changes in emotional experience.

\subsection{Data Collection}

\begin{figure}[!ht]
    \centering
    \includegraphics[width=0.75\linewidth]{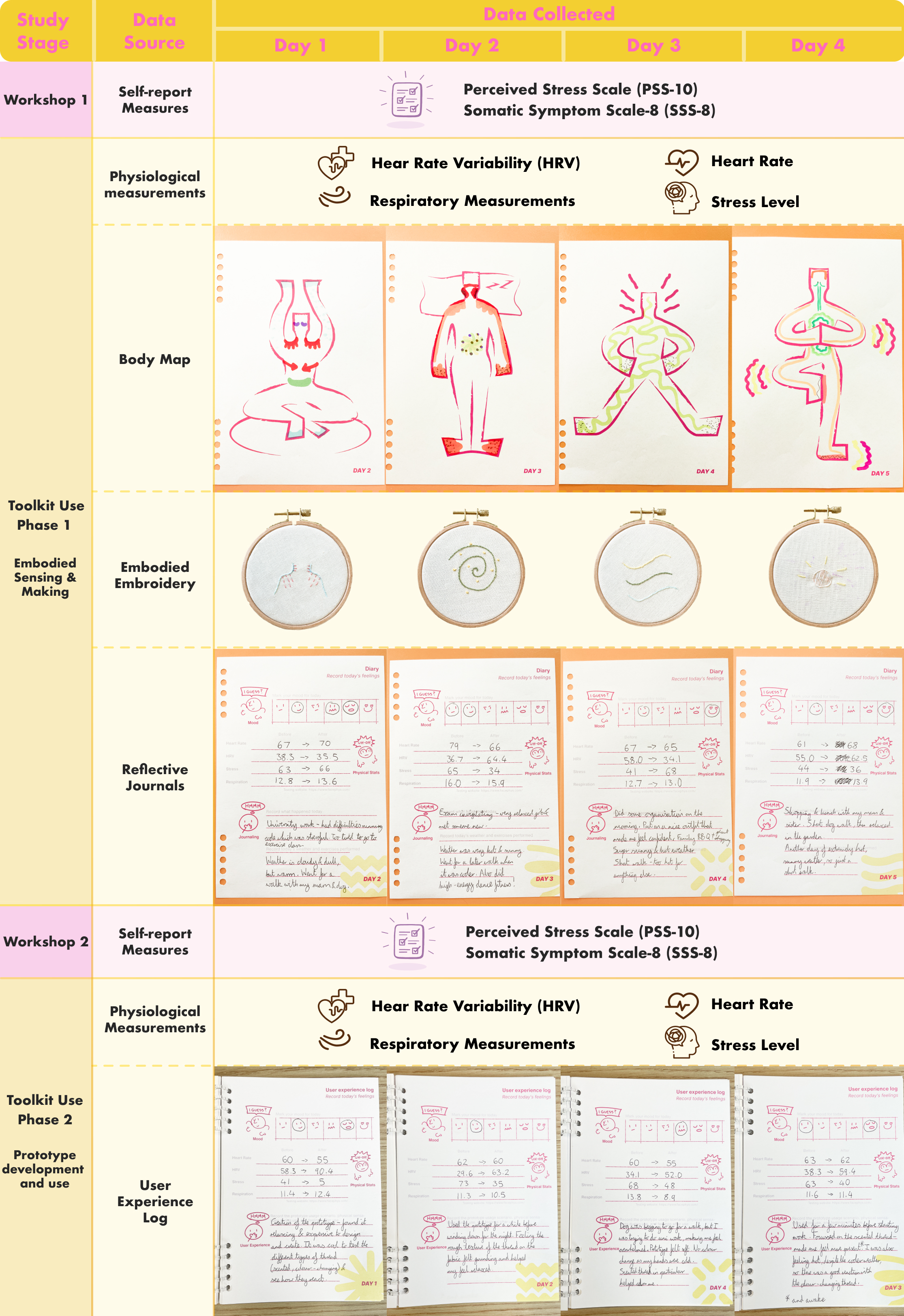}
    \caption{Qualitative and quantitative data matrix generated through a longitudinal diary and field study of embodied engagement.}
\Description{Matrix summarizing the qualitative and quantitative data collected across the longitudinal diary and field study. Workshop 1 and Workshop 2 included the Perceived Stress Scale (PSS-10) and Somatic Symptom Scale-8 (SSS-8). During Toolkit Use Phase 1, participants engaged with the toolkit in everyday settings while repeatedly recording heart rate variability, respiration, heart rate, and physiological stress, completing daily body maps and reflective journals, and producing embodied embroidery. During Toolkit Use Phase 2, participants developed and used their prototypes in everyday contexts while recording the same physiological measures and completing daily user-experience logs. The matrix brings together repeated quantitative measures, physiological data, embodied artifacts, body maps, journals, and experience logs across the two phases, with four example days illustrating the longitudinal structure of the dataset.}
    \label{fig:data}
\end{figure}

We collected multiple forms of data to examine participants' embodied engagement across different temporal and experiential scales: self-report measures, physiological data, reflective journals, semi-structured and micro-phenomenological interviews, and participant-made artifacts (see Figure~\ref{fig:data}).

\textbf{Self-report measures.}
Participants completed the Perceived Stress Scale (PSS-10)~\cite{cohen1983global} and the Somatic Symptom Scale--8 (SSS-8)~\cite{gierk2014somatic} at two time points: at the beginning of Workshop 1 and at the beginning of Workshop 2. The PSS-10 provided a measure of perceived stress, while the SSS-8 captured self-reported somatic symptoms. These measures were used to characterize changes in participants' perceived stress and somatic symptom experiences across the study, rather than to evaluate clinical or therapeutic efficacy of the toolkit.

\textbf{Physiological data.}
During the two five-day toolkit-use phases, participants recorded heart rate, HRV, stress level and respiratory measurements before and after selected activities, including embodied embroidery, and prototype use. Measurements were collected using a webcam-based, open-source facial physiology measurement platform.\footnote{\url{https://www.facephys.com/}} These measures provided a complementary view of participants' momentary physiological states during embodied engagement. Given that physiological signals can be influenced by movement, posture, environment, and other bodily factors, they were not treated as direct indicators of psychological stress. Instead, physiological data were considered alongside participants' accounts of their experiences during particular activities.

\textbf{Reflective journals.}
Participants used the paper notebook throughout the two toolkit-use phases to document their ongoing experiences. They were prompted to draw body maps and record what they noticed in their bodies and attention, as well as their subjective mood, weather conditions, and daily physical activity. These journal entries served as an important source of data for triangulating participants' self-reported experiences with physiological measurements, interview accounts, and participant-made artifacts. At the same time, they captured experiences close to when they occurred, providing insight into day-to-day variations and how participants independently adopted, adapted, or discontinued particular materials, activities, and practices over time.

\textbf{Interviews.}
The interviews generated two complementary forms of qualitative data: participants' retrospective accounts of their design processes, material choices, prototype development, and interpretations of bodily and emotional experiences; and detailed first-person accounts of specific moments of engagement with the toolkit. The latter captured fine-grained descriptions of bodily sensations, tactile and material experiences, attentional shifts, temporal unfolding, and emotional changes. Together, these interview data provided both broader reflective accounts of participants' engagement and detailed descriptions of particular embodied experiences.

\textbf{Participant-made artifacts.}
We documented the embodied embroidery pieces created during Workshop 1 and the prototypes developed following Workshop 2. These artifacts were treated as material traces of participants' interpretations, design decisions, and appropriations rather than simply as final outputs. Their forms, patterns, material combinations, and sensory or interactive features provided additional insight into how participants translated bodily experiences, reflections, and personal meanings into material configurations.

\subsection{Data Analysis}
Qualitative data were analyzed using reflexive thematic analysis \cite{braun2021one}, with attention to both patterned meanings across the dataset and the temporal unfolding of particular experiences. Following an iterative process of familiarization, coding, theme development, review, and refinement, we first examined the dataset across participants and data sources to identify recurring patterns, tensions, and divergences. We then returned to selected experiential episodes in greater detail, tracing how participants’ attention, bodily sensations, actions, and interpretations unfolded within particular moments of making and prototype use. This close analysis was informed by micro-phenomenological attention to the temporal and experiential qualities of concrete moments \cite{petitmenginDescribingOnesSubjective2006}, without treating it as a separate phenomenological analysis. This approach allowed us to retain cross-case patterns while examining how these patterns were differently constituted in situated experiences, drawing on recent methodological discussions of reflexive thematic analysis alongside phenomenologically oriented approaches \cite{phillips2026designing}. Additionally, quantitative data were used as a complementary source of evidence. We calculated descriptive statistics for the PSS-10 and SSS-8 and conducted within-participant pre–post comparisons. Physiological data were processed to derive participant-level measures of HR, HRV, and respiration, with measurements compared across relevant activity periods within each study phase. These analyses were used to identify patterns of change and variation that could be considered alongside participants’ qualitative accounts, rather than to establish causal effects or intervention efficacy.

Finally, we brought the different data sources into dialogue to identify convergences and divergences. We did not use triangulation to establish one definitive account of embodied experience. Instead, the different forms of data allowed us to examine embodied engagement across different temporalities and modes of articulation.

\subsection{Positionality and Reflexivity}

Our study was shaped by our positioning within embodied interaction, soma-oriented design, feminist HCI, and textile-based making. We approached embodiment as situated and plural rather than as a universal or stable capacity. This positioning shaped our attention toward differences and tensions in participants' experiences, rather than evaluating whether the toolkit produced a singular desirable outcome. Participants' decisions to follow, modify, combine, or reject particular materials and activities were consequently treated as meaningful forms of appropriation.

Our positionality also informs reflexivity throughout data analysis process. Rather than treating researcher subjectivity as a source of bias to be eliminated, we understood our interpretations as situated and actively produced through engagement with the data. Analytic decisions, emerging interpretations, tensions, and alternative readings were discussed iteratively within the research team. Our theoretical and design positioning therefore informed, but did not predetermine, the analysis: we remained attentive to experiences that complicated or challenged our expectations, including moments of distraction, discomfort, technical frustration, or disengagement. Such divergent and contradictory accounts were retained as part of the analysis rather than treated as exceptions to a dominant pattern.

\section{Findings}

\begin{figure}[!htbp]
    \centering
    \includegraphics[width=\linewidth]{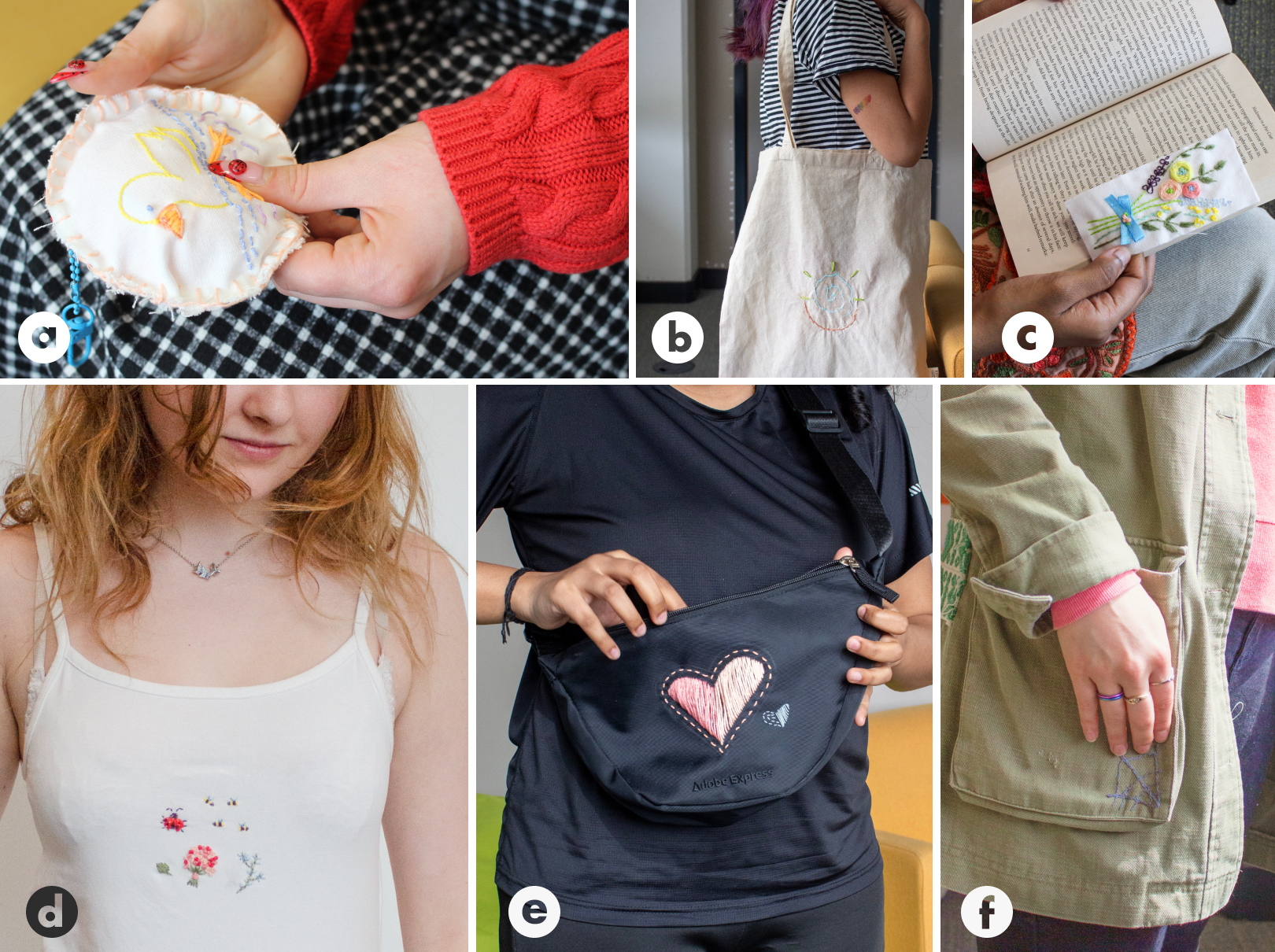}
    \caption{A collection of participants' artifacts: (a) padded duck bag charm by P8, (b) three-modality reminder tote bag by P7, (c) floral bookmark by P3, (d) summer scene tank top by P11, (e) luminous and color-changing heart bag charm by P9, and (f) spider web pocket by P4.}
   \Description{Six-panel collage of participants' embroidered artifacts. (a) A small padded duck-shaped bag charm with embroidered details. (b) A cream tote bag featuring a small multicolored spiral motif. (c) A floral bookmark decorated with embroidered flower motifs. (d) A light-coloured tank top featuring an embroidered summer scene with floral and outdoor elements. (e) A black bag incorporating luminous and color-changing hearts. (f) A coat pocket featuring a purple embroidered spider web motif.}
    \label{fig:findings}
\end{figure}

Overall, participants described the Embodied Embroidery Toolkit as a meaningful and often positive way of engaging with their bodily and emotional experiences. Across the study, participants reported feeling calmer, more focused, more grounded, or more aware of bodily sensations through embroidery, body mapping, and sensory engagement. The practice also provided ways of expressing experiences that were difficult to articulate verbally, while the material and technological possibilities encouraged participants to develop personally meaningful forms of reflection and self-regulation.

These experiences, however, did not emerge through a single or prescribed mode of engagement. Participants gradually developed their own ways of working with the activities, materials, and technologies according to their existing practices, bodily states, preferences, and everyday contexts. Variations in engagement therefore revealed not only when participants found the toolkit supportive, but also how they actively shaped the conditions of their own engagement. 

Across these accounts, we identify four interconnected aspects of participants' engagement: \textbf{negotiating structure and openness, sustaining attention through repetitive making, materializing stress through somatic anchors, and appropriating materials and technologies in situated ways}. Quantitative measures are incorporated where they provide complementary evidence alongside participants' accounts, rather than being treated as independent measures of intervention efficacy.

\subsection{From Guidance to Personal Ways of Engaging}
\subsubsection{Structure as an Entry Point}
Participants valued having some structure when entering and developing their embroidery practice, although they engaged with it in different ways. For some, planning reduced uncertainty and made the activity more manageable. P5 described how recognizing a stitching mistake prompted her to focus on the next step rather than becoming overwhelmed: \textit{``Once I knew I was doing it wrong, I came more focused on doing it right. So I have a plan what to do. I'm doing this wrong. Let's go on to the next thing and maybe feel a bit lighter, kind of happier.''} Similarly, P11 appreciated planning because it offered a more ``logical'' process: \textit{``And I like the embroidery. I like to like plan it out a bit. I guess more the logical side of it.''} She sketched designs beforehand and worked out how to distribute stitches as she progressed.

A sense of progression and completion provided another form of structure. P4 described the encouragement of seeing tangible progress: \textit{``That kind of inspired me to continue going on with the piece... And I think the pride in enjoying what I'm doing and physically being able to see measurements of progress was just a really nice feeling.''} For P3, completing individual elements was similarly important: \textit{``I don't like to leave things incomplete. So I try to think like, okay, so if I'm making this rose, I'm definitely going to sit down and complete the rose.''} Her completed bookmark became \textit{``proof that I can complete things that I've started...that I'm capable of sticking to one thing''}. Visible progress could therefore provide both a sense of accomplishment and evidence of participants' ability to sustain an activity.

\subsubsection{Developing Personal Structures}
As participants became more familiar with embroidery, structure was not only something provided by the toolkit but something they developed through their own routines. P6 described preparing the fabric, thread, needle, and threader as calming: \textit{``Even that in itself felt quite calming because it's like ritualistic. It's like every time you want to embroider, you have to use the same things and do the same process to get started.''} The repeated preparation thus became part of the embodied practice rather than simply a prerequisite for it.

Participants also developed ways of loosening the structure provided by instructions. P6 reframed mistakes as part of the making process: \textit{``It's not a competition or anything. There's nothing to get annoyed about... I started to enjoy even like the little shaky bits. I started to enjoy them really and look at them as part of the piece rather than like a mistake.''} Similarly, P7 initially sought precise guidance when working with e-textile materials, asking herself \textit{``what stitch should I do?''} and whether she was \textit{``supposed to''} use a particular technique. Later, she deliberately shifted away from the instructions: \textit{``I tried to figure it out by myself, like not so much focusing on the instructions as much. I feel like it was easier if I just went with it by myself.''} For her, following instructions too closely would have shifted attention toward performing the technique correctly rather than expressing her experience: \textit{``I would've been more focused on doing that correctly than just like portraying how I feel or what I felt.''}

These examples suggest that participants did not simply move from structured to unstructured engagement. Rather, they developed and adjusted structures according to what supported their engagement at a particular moment.

\subsubsection{When Structure Became an Obligation}
The supportive potential of structure was not unconditional. The same orientations toward planning and completion could become counterproductive when experienced as obligations. P1 described embroidery gradually becoming \textit{``a task that I have to complete''}, particularly as her patience decreased and she became preoccupied with when the piece would be finished. P9 similarly described how perfectionism made mistakes difficult to tolerate, leading her to repeatedly consider removing and redoing stitches.
These contrasting accounts show that structure could support engagement when it provided orientation without becoming an obligation to perform or complete. Participants benefited from ways of organizing their practice, but also needed room to change these structures when they no longer fitted their intentions or circumstances. Supportive structure was therefore not about providing more instructions or removing structure altogether, but about enabling participants to appropriate and recalibrate the degree of structure according to their own ways of engaging.

\subsection{Repetition as an Organizer of Attention}
\subsubsection{Repetition as an Attentional Scaffold}
Participants frequently described the repetitive movements of embroidery as ``calming'' (P6), ``focusing'' (P11), or ``meditative'' (P3). For several participants, repetitive stitching helped organize attention around the immediate activity by reducing the need to continually decide what to do next.

P3 described making roses as a highly repetitive process. Once she entered what she called the ``groove'', the repetition became meditative:

\begin{quote}
``But once I got into the groove of it, cuz making the roses is a pretty repetitive process. [...] I was just in this meditation in itself. I had done the body scan meditation before that, but during the process of embroidery itself, I felt like I was meditating.''
\end{quote}

Similarly, P11 described the repetitive movement as a form of ``focused relaxing'', explaining that \textit{``by time you get into that repeated motion, you're just like doing it over and over.''} For P7, repetition eventually reduced the cognitive demand of the activity and created space for self-attention: \textit{``Once it starts making sense it's just repetitive movements, which is nice because it allows you to not be so focus on the activity itself, but also yourself.''} Here, repetition supported attention partly by making the activity more predictable and less cognitively demanding.

\subsubsection{Anchoring Attention in Sensory-Motor Engagement}
Repetition also directed attention toward the sensory and motor details of making. P6 described deliberately focusing on her hands and the material, attending to the feeling of the thread passing through the fabric, pushing the needle, and tying knots. During one session, this concentration became sufficiently absorbing that she noticed she had stopped attending to the music playing in the background: \textit{``sometimes I'd like finish a stitch and then be, oh, I didn't even realize that song was on because I was just so focused on it.''} She described her hands, fingers, and fabric as becoming the most important things in her immediate environment. This focus could also extend to everyday settings. While embroidering on a train, P6 became less concerned with how others might perceive her because she was \textit{``just focused on my stitches''}.

For P4, the relationship between repetition and attention was particularly evident in her chain-stitch spider web. She deliberately chose the more time-consuming stitch because its repetitive and precise movements required sustained attention. She described the process as self-soothing, explaining that focusing on keeping the stitches neat helped her attend to the movement rather than other thoughts:

\begin{quote}
``I use that kind of repetitive motion as like a self soothing because it's repetitive and it's focus on doing it, keeping it neat, getting it nice and neat, making it look good that having that focus help. To calm me down whilst doing the embroidery.''
\end{quote}

%Attention was not limited to the stitching process. During body scanning, P10 deliberately moved attention from her right arm to her left arm and through her fingers, describing the experience as `really fun and joyful'' and `very distinct'' for her. Here, shifting attention across bodily locations became an experiential activity in its own right. Her journal, however, also included moments of numbness, irritation, and fatigue, suggesting that bodily attention varied across situations.

Together, the discussions demonstrate how repetitive and sensory-motor engagement provided concrete anchors for attention. Rather than requiring participants to maintain an abstract intention to ``be mindful'', stitching offered specific movements, sensations, and material details to notice and attend to.

\subsubsection{When Repetition Did Not Redirect Attention}

Repetitive movement did not automatically produce embodied attention. The same motor rhythm could also allow participants' attention to wander or remain occupied by concerns outside the activity. P6, for example, described repetition as something that could allow her mind to wander, indicating that reduced cognitive demand did not necessarily mean sustained attention to the body or material.
This became particularly consequential in P2's experience. She reported that during one session she was physically engaged in embroidery while her thoughts remained occupied by daily stressors:

\begin{quote}
It is very important to be mindful because there is one such instance where my stress levels increased after the embroidery session that day. So I think even while doing embroidery, if our mind keeps waving here and there thinking about, you know, the daily stress and other activities. Then I guess it's will not be of much use. It's important to practice mindfulness as well.
\end{quote}

Her individual measures provided a contrasting pattern. Her PSS-10 increased from 27 to 31 and her SSS-8 from 18 to 20 across the study. During Phase 1 Day 4, her physiological stress score increased from 28 to 67 after embroidery, while her heart rate decreased from 82 to 74 bpm and HRV increased from 63.0 to 69.9 ms. These measures illustrate that physiological indicators did not move in a single direction and should not be collapsed into a simple account of stress reduction. In P2's case, stitching continued while her attention remained entangled with other concerns.

In contrast, P6 showed a more convergent pattern across experiential, self-report, and physiological traces. During Phase 1 Day 3, her HRV increased from 54.0 to 121.5 ms and her recorded physiological stress score decreased from 46 to 0. Her PSS-10 and SSS-8 scores also decreased from 27 to 13 and from 18 to 11, respectively. These changes accompanied her account of becoming absorbed in the repetitive and sensory aspects of making. We do not treat these measures as evidence that repetition caused stress reduction, but as complementary traces that coincided with her reported experience of sustained attentional engagement.

Overall, the value of repetition lay less in repetition itself than in the attentional relationship it enabled. Repetitive stitching could narrow attention onto movement and material, provide a steady rhythm for entering a flow-like state, or reduce the cognitive demands of making enough to create space for self-attention. It could also coexist with mind-wandering or continued mental preoccupation. Embodied attention was therefore not a stable state produced by repetition, but something participants entered, lost, and reconfigured through their engagement with the practice.

\subsection{Materializing Stress Through Somatic Anchors}
\subsubsection{Tactility as an Invitation to Engage}
Participants used the toolkit not only to notice bodily states during the activity, but also to translate these experiences into tangible forms that could be revisited beyond the moment of reflection. Tactile qualities were particularly important in creating opportunities for repeated bodily engagement.

P4 experimented with different thread weights and stitch types to create contrasting tactile sensations. She combined four and two strands with French knot stitches to create changes in texture across a spider-web motif, which she placed on the pocket of her everyday jacket where her dominant hand could reach it easily. She described the result as ``really tactile'' and explained:

\begin{quote}
``And then I was running my finger over one where I'd played with the thread weight. And then realized it was very tactile and that I liked running my finger over in like a self soothing way. So on the pocket of my jacket, on my dominant hand, is a tiny little spider web. [...] at the ends of the three kind of outward lines is a French knot stitch, so there's a little bit more tactile when you've reached the edge.''
\end{quote}

The tactile surface subsequently became part of how P4 engaged with her everyday routine. She imagined rubbing or fiddling with it when her usual routine was disrupted, describing it as ``my little self soothing kind of rubbing, fiddling, kind of tactile ness''. The artifact thus became a portable tactile resource incorporated into an existing bodily habit.

P8 similarly developed a portable artifact using soft filling, thicker colour-changing thread, and scented thread. She described touching and squeezing the thicker thread, smelling the scented thread when worried, and watching the colours change:

\begin{quote}
Cuz the color changing threads are a lot thicker, so I just find myself feeling the texture of them. And then like squishing or if I'm feeling a bit more worried, then I'll smell the center thread as well. But I just like to watch the colors change.
\end{quote}

These different sensory actions gave P8 several ways of engaging with the artifact during everyday moments of concern. On one evening when she felt overwhelmed, she used the prototype and described the scented thread as particularly calming. Her PSS-10 decreased from 17 to 13 and her SSS-8 from 2 to 1 between the two study time points. During Phase 2 Day 1, her heart rate decreased from 85 to 71 bpm, HRV increased from 46.2 to 60.0 ms, and recorded physiological stress decreased from 54 to 36. These measures provide complementary traces alongside her account rather than evidence of a causal effect.

\subsubsection{Materializing and Revisiting Bodily Experience}

Material engagement also allowed participants to give form to experiences that were difficult to articulate. P7 described a spiral motif as representing the experience of gradually ``untangling'' her body and senses through body scanning, meditation, and embroidery. She connected the process with becoming better able to understand and express feelings that had previously been difficult to make sense of. The artifact therefore captured an unfolding process of noticing and interpretation rather than simply representing a fixed state of stress.

Scent provided another way of extending this process beyond the original activity. P7 selected scented thread partly because she wanted it to function as ``a nice reminder''. Together with the luminous and color-changing threads, the scent contributed to the meaning of her piece and its relaxing quality. Unlike a written instruction, it could be encountered through a brief sensory interaction in everyday life, allowing the artifact to evoke the practice without requiring a separate reflective activity.

Materialization was also connected to a sense of visible progress and achievement. P4 described looking down at her embroidery and realizing that she had completed more than expected:

\begin{quote}
``That then I looked down and was like, oh, I've done more than I thought I've done. And now this is going to look good. And I'm starting to get a formalized idea of what the bigger picture will be. [...] the pride in enjoying what I'm doing and physically being able to see measurements of progress was just a really nice feeling.''
\end{quote}

P5 similarly described looking down at her personalized tote and seeing that she had added her own design to it: \textit{``It's kind of cool. I added to it and took my own design upon it. So yeah, it's nice looking down and seeing that I've done something on it.''} P11 also emphasized the significance of being able to see and recognize what she had made. The completed artifact thus made progress physically visible and provided a material point of reference for participants' sense of accomplishment.

For P4, this change was also reflected in her self-report measures. Her PSS-10 decreased from 34 to 24 and her SSS-8 from 15 to 13. Across Phase 1, her heart rate remained relatively low and stable, while recorded physiological stress decreased from 71 on Day 1 to 50 on Days 2 and 3. These traces are consistent with her accounts of increasing enjoyment and visible progress, but do not establish what produced the change. Her experiential accounts remain important for understanding how making and material engagement contributed to this experience.

Materialization could also extend to identity. P6 described feeling proud of creating something wearable and explained that seeing her idea become tangible made her feel \textit{``more in touch with my identity because this is something that I just came up with.''} The wearable artifact thus connected creative agency with how she presented and experienced herself.

\subsubsection{Somatic Anchors Across Time and Context}

Across these cases, somatic anchors developed through an ongoing process rather than existing as fixed features of the toolkit. Participants noticed bodily experiences, experimented with materials, shaped sensory and visual qualities according to what felt meaningful, and then carried the resulting forms into subsequent situations. An anchor could therefore be touched, smelled, viewed, worn, or carried after the original making activity.

This process connected different temporal dimensions of embodied engagement. Bodily noticing occurred during particular moments of reflection, while material experimentation transformed those experiences during making. The resulting artifacts could then provide further points of bodily and sensory engagement in everyday life. What was materialized was therefore not simply ``stress'', but participants' interpretations of bodily experience and the ways they wanted to relate to it.

Across the study, self-report measures, physiological traces, artifacts, and first-person accounts offered different forms of evidence about these experiences. Their value lay not in establishing a single objective account of stress, but in showing how changes in experience could be understood alongside the material practices through which participants noticed, expressed, and revisited their bodily states.

Overall, somatic anchors made bodily experience tangible enough to notice and revisit. Through making, participants developed tactile, sensory, visual, and wearable forms that could subsequently support attention, self-soothing, reflection, a sense of accomplishment, and personal meaning in everyday contexts.

\subsection{Situated Appropriation of Materials and Technology}

Participants did not uniformly explore the toolkit's materials and technologies. Instead, they selected, combined, modified, or reconfigured particular features according to their everyday practices, disciplinary orientations, bodily conditions, and intended uses. Technology was valued not simply for its technical complexity, but for how it became meaningful within participants' own ways of engaging with the practice.

\subsubsection{Appropriating for Everyday Use}

Several participants adapted the toolkit towards forms that could fit into their everyday lives. P8, for example, designed a portable bag charm that could be carried with her and used when she felt stressed. She chose thermochromic thread that responded to body heat and combined it with scented thread:
\textit{``I used a mixture of the color changing, but I kept it analog. So it just reacts to body heat. And mainly just so that it's portable.''}
Rather than maximizing the prototype's technological features, P8 retained the feature that best fit her intended use. Similarly, P1 emphasized subtlety when designing an embroidered T-shirt: \textit{``I just wanted to keep it [embroidered T-shirt] subtle ... match with my everyday life.''} For her, the textile was intended to become an everyday reminder rather than an attention-demanding interactive object.

P6 adapted the practice to a less predictable everyday setting. While embroidering on a train, she initially found the movement frustrating but gradually incorporated it into her technique:

\begin{quote}
``Initially, I was pretty frustrated because it [the train] was like shaking a lot. But then instead of getting more annoyed, I tried to lean into that and like enjoy how precise it [the embroidered artifact] needed to be. [...] I feel like it just made the piece more enjoyable because it was like sort of unintentionally incorporating the environment as well.''
\end{quote}

Here, the environment was not simply where embroidery took place, but also part of how the activity was performed and experienced.

\subsubsection{Negotiating Material and Bodily Conditions}

Appropriation also emerged through participants' encounters with material properties and bodily conditions. P9 was particularly curious about the color-changing and luminous threads, but found them physically demanding to use. Her diary described the materials as difficult to embroider, while sweaty hands further increased the difficulty:

\begin{quote}
``It's already hard to put the needle through this material, but with that thick thread, I was actually like using my shirt to like pull it through like this. But also like I said, my hands get sweaty very fast, so it was a struggle.''
\end{quote}

The same material could therefore be experienced differently depending on the body encountering it. For P9, its interactive properties made it interesting, while its physical properties simultaneously created difficulties.

Bodily circumstances could also affect how manageable the overall practice became. P9 described becoming overwhelmed when tangled threads and limited material availability coincided with being on her period:

\begin{quote}
``These two were supposed to be separate. It got all tangled up in the back that I could not make sense. And there was a big loop and this lot of thread wasted and I did not have enough to complete. Then I think I just screamed or something and I was on my periods at that point, so I started crying. And my partner was like, isn't this supposed to make you calm? Why are you crying?''
\end{quote}

This episode contrasts with accounts of embroidery as calming or absorbing, showing that the practice did not encounter a singular or standardized body. What was manageable in one situation could become difficult or overwhelming in another.

At the same time, material resistance did not necessarily lead participants to reject particular features. P7 found the luminous thread particularly rewarding because she had never encountered it before, describing learning to use it as \textit{``more rewarding than if I had just used like a normal thread''}. The interactive properties made the prototype more personal and playful for her. Participants thus engaged with materials selectively, retaining, modifying, or working around properties according to what they found meaningful or workable.

\subsubsection{Reconfiguring Technology Through Disciplinary Orientations}

Participants' disciplinary experiences also shaped how they approached the relationship between technology and making. P7 described coding as involving \textit{``a lot of...logic''}, where there was generally \textit{``only one way to get it right''}. She initially brought this orientation into embroidery by trying to follow instructions closely and make her work perfect. Through the activity, however, she came to value a more flexible mode of engagement: \textit{``you just need to do what feels right for you and there's no right and wrong.''} Embroidery thus offered a different relationship to making in which correctness was less clearly defined.

This did not mean that participants simply moved from technology towards craft. P7 remained interested in the technical properties of e-textiles, while P8 deliberately simplified her prototype by retaining only the feature that fitted her intended use. For P6, P8, and P11, embroidery also offered a different way of engaging with making beyond their usual technical practices.

Across these cases, appropriation was part of how embodied engagement was produced. Participants decided what to use, how to use it, what to modify, and what role technology should play. Situated appropriation, therefore, emerged through negotiation among materials, bodies, disciplinary orientations, and everyday contexts.

\section{Discussion}

%repetitive stitching could help narrow attention onto the hands and materials, create a steady rhythm, or allow the mind to wander. These different forms of attention shaped whether repetition was experienced as soothing, distracting, or even stressful.
%Embroidery therefore did not inherently produce mindfulness or calmness; rather, \textbf{its repetitive and sensory qualities created possibilities for different modes of attention. Designing for embodied engagement consequently requires attention not only to the motor characteristics of an activity, but also to how participants are supported in directing, shifting, or releasing their attention during making. 

\subsection{Embodied Engagement is Negotiated}
Embodied interaction has long challenged views of interaction as primarily cognitive or representational, emphasizing instead how interaction emerges through bodily action situated in everyday contexts \citep{dourish2001where}. Soma design extends this orientation by treating first-person bodily experience as a source of knowledge and transformation, and by seeking to cultivate sensitivity to subtle bodily experiences \citep{hook2018soma}. In these approaches, design often aims to create conditions that invite people to notice, attend to, and make sense of their bodily experiences. Our findings extend this perspective by showing that embodied engagement unfolds dynamically over time, as participants' attention and sense of involvement shift in relation to the structure of the activity and their immediate circumstances.

This temporal quality was particularly evident in participants' relationships with structure. Instructions, examples, and progressive tasks helped some participants enter an unfamiliar practice, while planning and visible milestones provided others with a sense of orientation and accomplishment. At the same time, these structures could become experienced as obligations when participants became concerned with completing a task or performing a technique correctly. Structure shaped the conditions under which engagement could emerge, but did not determine the quality of that engagement. Participants moved between moments of focused involvement, task-oriented attention, and disengagement as they worked through the practice.

Attention was similarly dynamic. Repetitive embroidery could create a steady rhythm, reduce cognitive demands, and support attention to hands, materials, and movement. Participants described these experiences as calming, meditative, or grounding. At other moments, however, repetition allowed attention to wander or remain occupied by unrelated concerns. P2's experience illustrates this tension particularly clearly: although she continued to stitch, her attention remained caught up in everyday stress, and she described an increase rather than a reduction in stress during one episode. P4, by contrast, deliberately used the precision and repetition of chain stitching to focus her attention and quiet racing thoughts. The same motor activity could therefore support different forms of engagement depending on how attention was organized within the moment.

The individual quantitative traces further support this situated and temporal interpretation. For P6, physiological measures, self-report measures, and experiential accounts broadly converged around moments of absorbed and calming engagement. For P2, these traces were more complex, with physiological indicators and self-reported stress not always moving in the same direction, while her lived account emphasized continued mental preoccupation. Rather than resolving these differences by privileging one source of data, we treat them as evidence of the multiple temporal and experiential dimensions through which embodied engagement becomes visible. What appears as calmness in one measure may coexist with cognitive preoccupation in another, and these states may change across a single making episode.

Our findings suggest that designing for embodiment is not simply a matter of selecting activities assumed to produce bodily awareness or calmness. Instead, designers can create conditions that support different modes of engagement and allow attention to shift throughout an activity. This involves considering how structure, repetition, pacing, and opportunities for pause shape participants' capacity to enter, sustain, redirect, or disengage from an embodied practice. Embodied engagement can therefore be understood as an unfolding process rather than a stable outcome produced by a particular activity or material.

\subsection{Making Space for Plural Embodied Experiences}
Embodied and soma-oriented approaches have positioned the lived body as central to perception, action, and meaning-making \citep{hook2018soma}, while feminist HCI has challenged assumptions of a universal user by attending to how gender, identity, social context, and power shape interaction \citep{10.1145/3196839.3196845}. Work on women's health has similarly foregrounded situated bodily experience rather than treating the body solely as an object of measurement or intervention \citep{park2025designing, hua2023embodied}. Our findings extend this orientation by showing that plurality in embodied experience is also produced through the relationship between particular bodies, material properties, and situated circumstances.

Participants encountered the same materials through markedly different bodily conditions. P9 found the stiffness of luminous and color-changing threads difficult to manage, particularly when heat and sweaty hands increased the physical effort required. In contrast, P5 and P7 experienced interactive threads as novel, rewarding, or engaging. These differences cannot be understood simply as differences in individual preference. They emerged through the interaction between material properties and particular bodily circumstances. Material affordances were therefore not universal properties of the toolkit, but became meaningful through specific bodies and contexts.

This plurality also appeared in how participants incorporated movement and tactile sensation into making. P3 described herself as fidgety and frequently changed body positions while working, while P4 incorporated tactile fidgeting into her everyday routine and deliberately varied thread weights to create different sensations. P10 reconfigured the response of the color-changing thread into repeated rubbing against her joints. These practices complicate the idea that embodied engagement requires a particular posture, attentional state, or form of bodily awareness. Participants instead developed different relationships with movement, touch, sensation, and repetition according to what was meaningful or comfortable for them. Consequently, embodied participation took multiple forms within the same designed practice.

The findings also complicate a simple distinction between accessible and inaccessible materials. Material resistance could function as a barrier, as in P9's experience, but could also become a meaningful part of the artifact. P4, for example, incorporated exposed knots and variations in thread weight into the tactile quality of her piece. Material friction was therefore not necessarily something to be eliminated. Its significance depended on how participants experienced and interpreted it within a particular activity.

This perspective shifts accessibility from designing for a normative embodied user towards making space for different embodied relations. Participants brought different sensory preferences, bodily conditions, routines, motor habits, and identities into the same practice, and these differences shaped what the materials could afford. Designing embodied interactions should therefore account for bodily difference not only as a constraint, but also as a source of variation in how interactions become meaningful. Rather than prescribing a single appropriate way of sensing, moving, or making, embodied design can provide materials and activities that remain open to adjustment, interpretation, and different sensory pathways.

\subsection{From Embodied Technology to Embodied Appropriation}
HCI has long recognized that technologies are not used exactly as anticipated by designers. Technology appropriation describes how people incorporate, reinterpret, and adapt artifacts within their situated practices rather than simply following designers' intended uses \citep{dix2007designing, tchounikine2017designing}. Such appropriation is also shaped by social and identity contexts. Rode and Poole, for example, show that engagement with digital technologies is intertwined with gender identity and technical identity, challenging accounts that explain technology engagement primarily through technical expertise \citep{10.1145/3196839.3196845}. Building on this work, our findings show how appropriation in embodied design can involve not only the functional use of technology, but also the selective incorporation of technological properties into participants' bodily, material, and everyday practices.

Participants did not need to engage with the toolkit's full technological capabilities for it to become meaningful. P8, for example, retained the thermochromic thread in a largely analog bag charm, valuing its responsiveness to body heat while keeping the artifact portable and simple enough to incorporate into everyday life. Her selective use of the technology was not a reduction of technological engagement, but a way of identifying and retaining the property that mattered to her. The technological feature became meaningful precisely because it could be incorporated into an existing practice and personal object.

Appropriation could also involve reconfiguring the relationship between the material and the surrounding environment. P6 initially experienced difficulty embroidering on a moving train, but gradually adapted her stitching and incorporated the instability of the environment into the making process. Here, the train's changing movement became part of the conditions of making rather than simply an obstacle to be overcome. The resulting practice was shaped jointly by the participant, the material, and the environment, illustrating how appropriation can extend beyond the artifact itself to include the circumstances in which it is used.

Appropriation was further shaped by participants' existing disciplinary practices and identities. For participants accustomed to computational work, embroidery sometimes offered an alternative relationship to making. P7 contrasted the open-endedness of embroidery with the logic of programming, where there could be a more clearly defined correct solution. At the same time, participants did not simply leave their technical orientations behind. Some remained curious about the technical properties of e-textiles, while others valued them primarily for their expressive or sensory qualities. The toolkit consequently became a space where established ways of making and thinking could coexist with, and sometimes be reoriented through, a different material practice.

The examples suggest that appropriation was integral to how the embodied practice developed. Participants gave the toolkit personal relevance by selecting particular properties, incorporating them into existing practices, and reinterpreting them through their own technical and material orientations. This extends conventional accounts of technology appropriation by highlighting its embodied dimension: what becomes appropriated is not only a technological function, but also a particular way of sensing, making, moving, and relating to an artifact. Embodied appropriation thus describes a process through which technological possibilities become situated within participants' bodies, practices, and everyday lives.

\subsection{Limitations \& Future Work}
Although we use STEM as the broader disciplinary category, our participants were predominantly situated in computing and computational disciplines. Their experiences may therefore reflect particular cultures of computational education, including programming, technical problem-solving, and screen-based work. This disciplinary concentration limits the extent to which our findings can be generalized across STEM contexts. Future work could compare embodied engagement across different STEM disciplines, particularly those involving different relationships with bodies, materials, laboratories, and computational practice.

The study was conducted across two countries [anonymized for review], but the present paper does not undertake a systematic cross-cultural analysis. This limits our ability to examine how cultural expectations, educational environments, and craft traditions may shape participants' embodied engagement. Future comparative analysis could examine which aspects of embodied engagement recur across contexts and which are shaped by culturally specific understandings of gender, well-being, craft, and academic practice. Such analysis could further clarify how embodied design practices are situated within different educational and cultural environments.

\section{Design Sensibilities for Negotiating Embodied Engagement}
Our study suggests that designing for embodied engagement involves creating conditions that participants can adapt to their own ways of sensing, attending, making, and using. Rather than prescribing a particular embodied experience, we articulate four design sensibilities for supporting such adaptation.

\subsection{Scaffold without prescribing}
Participants appreciated having a clear starting point, particularly when embroidery or reflective activities were unfamiliar. At the same time, closely following instructions could turn the activity into another task to complete. We therefore suggest \textbf{providing structure as an invitation rather than a fixed route}. Designers can offer accessible entry points, examples, and optional guidance while leaving participants room to change techniques, skip activities, establish their own routines, or reinterpret the intended use of materials. This allows structure to reduce uncertainty without becoming another form of performance.

\subsection{Support multiple ways of attending}
Embodied engagement took different forms across participants. Some focused closely on the movement of stitching and the qualities of the material, while others shifted attention between bodily sensations, the activity, and their surrounding environment. Repetition could support concentration, reflection, or moments in which the mind was allowed to wander. We therefore suggest \textbf{supporting multiple ways of attending rather than designing toward a single attentional state}. Embodied interactions can provide opportunities to focus, shift, linger, or return attention without requiring participants to remain continuously mindful or treating distraction as failure.

\subsection{Make embodied experience tangible and revisitable}
Participants turned bodily and emotional experiences into artifacts that could be touched, viewed, smelled, worn, and carried beyond the immediate activity. Tactile surfaces could become objects for fidgeting or grounding, scent could serve as a subtle reminder, and completed artifacts could provide visible evidence of progress and personal meaning. We therefore suggest \textbf{creating material anchors that make embodied experience tangible and revisitable}. Designing for portability, wearability, and repeated sensory access can allow an embodied practice to extend into everyday life rather than ending with the original interaction.

\subsection{Leave room for situated appropriation}
Participants adapted materials and technologies according to their bodies, interests, disciplinary habits, and circumstances. Some kept interactions analog or subtle, while others valued the novelty of e-textiles. Material difficulties could also lead participants to modify their techniques or incorporate unexpected properties into the finished artifact. We therefore suggest \textbf{treating materials and technologies as resources for appropriation rather than fixed interaction mechanisms}. Optional, reconfigurable components can allow participants to decide how much technical, sensory, and physical complexity fits their own needs and contexts.

\section{Conclusion}
We examined how female STEM students engaged with and appropriated an Embodied Embroidery Toolkit designed to support exploration of bodily and emotional experiences. Across two five-day toolkit-use phases, participants developed diverse ways of engaging with embroidery, sensory materials, body mapping, and optional e-textiles. Rather than producing a uniform embodied response, the toolkit became a space in which participants negotiated structure, attention, bodily sensations, materials, technologies, and everyday contexts.

We conceptualize this process as embodied engagement as situated negotiation. Participants did not simply follow the practices and materials provided, but adapted them according to their bodies, identities, disciplinary orientations, and circumstances. Repetition could organize attention while also allowing distraction; materials could make bodily experiences tangible while generating different forms of friction; and technologies could be retained, simplified, or reconfigured according to situated needs.

Based on these findings, we identify four design sensibilities for supporting embodied engagement: scaffolding without prescribing, supporting multiple ways of attending, making embodied experiences tangible and revisitable, and leaving room for situated appropriation. Together, these sensibilities suggest a shift from designing for particular embodied outcomes towards designing conditions in which people can develop their own ways of engaging with bodily experience.

\bibliographystyle{ACM-Reference-Format}
\bibliography{reference}

\appendix

\section{Declaration of Generative AI Use} 
Generative AI (GenAI) tools were not used to create the research questions, design the study, produce workshop artifacts, or analyze the data. GenAI was used only during the final stages of manuscript preparation to assist with language editing and the refinement of grammar and style. The authors reviewed and revised all AI-assisted text and remain fully responsible for the accuracy and integrity of the manuscript, ensuring that it contains no fabricated or plagiarized material.

\section{Quantitative Data}
\label{app:quantitative_data}

\subsection{Somatic Symptom Scale 8}
\label{app:sss8}

\subsubsection{SSS-8 Items and Response Scale}
\label{app:sss8_items}

\begin{table}[!htbp]
\centering
\caption{SSS-8 item descriptions and response scale.}
\label{tab:sss8_items}
\small

\begin{tabular}{cl}
\toprule
Item & Description \\
\midrule
Q1 & Stomach or bowel problems \\
Q2 & Back pain \\
Q3 & Pain in your arms, legs, or joints \\
Q4 & Headaches \\
Q5 & Chest pain or shortness of breath \\
Q6 & Dizziness \\
Q7 & Feeling tired or having low energy \\
Q8 & Trouble sleeping \\
\bottomrule
\end{tabular}

\vspace{2mm}

\begin{minipage}{0.90\textwidth}[!htbp]
\footnotesize
\textit{Response scale:}
0 = Not at all;
1 = A little bit;
2 = Somewhat;
3 = Quite a bit;
4 = Very much.
\end{minipage}

\end{table}

% ------------------------------------------------------------
% B.2 SSS-8 Participant Scores
% ------------------------------------------------------------

\subsubsection{SSS-8 Participant Responses}
\label{app:sss8_scores}

\begin{table}[!htbp]
\centering
\caption{Participant SSS-8 responses in Phase 1 and Phase 2.}
\label{tab:sss8_scores}
\scriptsize
\setlength{\tabcolsep}{5pt}

\begin{tabular}{llrrrrrrrrr}
\toprule
Phase & Participant
& Q1 & Q2 & Q3 & Q4 & Q5 & Q6 & Q7 & Q8 & Sum \\
\midrule

\multicolumn{11}{l}{\textit{Phase 1}} \\

P1  & & / & 4 & 1 & / & 2 & 4 & 2 & 4 & 17 \\
P2  & & 1 & 3 & 1 & 4 & 3 & 0 & 2 & 4 & 18 \\
P3  & & 0 & 2 & 3 & 4 & 0 & 2 & 4 & 4 & 19 \\
P4  & & 0 & 1 & 2 & 2 & 0 & 3 & 4 & 3 & 15 \\
P5  & & 2 & 4 & 1 & 4 & 3 & 0 & 1 & 2 & 17 \\
P6  & & 4 & 2 & 3 & 0 & 0 & 1 & 4 & 4 & 18 \\
P7  & & 3 & 3 & 3 & 2 & 2 & 1 & 3 & 3 & 20 \\
P8  & & 1 & 0 & 1 & 0 & 0 & 0 & 0 & 0 & 2 \\
P9  & & 3 & 2 & 1 & 4 & 0 & 1 & 3 & 0 & 14 \\
P10 & & 0 & 2 & 0 & 0 & 2 & 0 & 1 & 0 & 5 \\
P11 & & 0 & 0 & 0 & 1 & 0 & 0 & 1 & 2 & 4 \\

\midrule

\multicolumn{11}{l}{\textit{Phase 2}} \\

P1  & & 1 & 4 & 3 & 0 & 2 & 0 & 2 & 0 & 12 \\
P2  & & 1 & 3 & 2 & 4 & 3 & 0 & 4 & 3 & 20 \\
P3  & & 0 & 2 & 0 & 3 & 0 & 2 & 3 & 4 & 14 \\
P4  & & 0 & 2 & 3 & 0 & / & 2 & 3 & 3 & 13 \\
P5  & & 1 & 2 & 3 & 2 & 2 & 1 & 0 & 1 & 12 \\
P6  & & 4 & 2 & 1 & 0 & 0 & 1 & 2 & 1 & 11 \\
P7  & & 2 & 1 & 3 & 1 & 1 & 1 & 2 & 0 & 11 \\
P8  & & 0 & 0 & 1 & 0 & 0 & 0 & 0 & 0 & 1 \\
P9  & & 4 & 4 & 3 & 3 & 0 & 0 & 3 & 4 & 21 \\
P10 & & 1 & 0 & 3 & 0 & 1 & 0 & 3 & 1 & 9 \\
P11 & & 0 & 0 & 1 & 0 & 0 & 0 & 2 & 1 & 4 \\

\bottomrule
\end{tabular}

\vspace{2mm}

\begin{minipage}{0.90\textwidth}[!htbp]
\footnotesize
\textit{Note.}
``/'' indicates that the participant omitted the item.
Higher total scores indicate greater somatic symptom burden.
\end{minipage}

\end{table}

% ============================================================
% C. PSS-10
% ============================================================

\subsection{Perceived Stress Scale 10}
\label{app:pss10}

% ------------------------------------------------------------
% C.1 PSS-10 Items
% ------------------------------------------------------------

\subsubsection{PSS-10 Items and Scoring}
\label{app:pss10_items}

\begin{table}[!htbp]
\centering
\caption{PSS-10 item descriptions and scoring direction.}
\label{tab:pss10_items}
\scriptsize

\begin{tabular}{cl}
\toprule
Item & Description \\
\midrule
Q1 & Been upset because of something that happened unexpectedly? \\
Q2 & Felt that you were unable to control the important things in your life? \\
Q3 & Felt nervous and ``stressed''? \\
Q4 & Felt confident about your ability to handle your personal problems? \\
Q5 & Felt that things were going your way? \\
Q6 & Found that you could not cope with all the things that you had to do? \\
Q7 & Been able to control irritations in your life? \\
Q8 & Felt that you were on top of things? \\
Q9 & Been angered because of things that were outside of your control? \\
Q10 & Felt difficulties were piling up so high that you could not overcome them? \\
\bottomrule
\end{tabular}

\vspace{2mm}

\begin{minipage}{0.90\textwidth}
\footnotesize
\textit{Response scale:}
0 = Never;
1 = Almost never;
2 = Sometimes;
3 = Fairly often;
4 = Very often.

Items Q4, Q5, Q7, and Q8 are reverse-scored.
Higher total scores indicate higher perceived stress.
\end{minipage}

\end{table}

% ------------------------------------------------------------
% C.2 PSS-10 Participant Scores
% ------------------------------------------------------------

\subsubsection{PSS-10 Participant Responses}
\label{app:pss10_scores}

\begin{table}[!htbp]
\centering
\caption{Participant PSS-10 responses in Phase 1 and Phase 2.}
\label{tab:pss10_scores}
\scriptsize
\setlength{\tabcolsep}{4pt}

\begin{tabular}{llrrrrrrrrrrr}
\toprule
Phase & Participant
& Q1 & Q2 & Q3 & Q4 & Q5 & Q6 & Q7 & Q8 & Q9 & Q10 & Sum \\
\midrule

\multicolumn{12}{l}{\textit{Phase 1}} \\

P1  & & 4 & 3 & 4 & 2 & 2 & 4 & 2 & 3 & 3 & 4 & 31 \\
P2  & & 3 & 3 & 4 & 2 & 3 & 3 & 2 & 2 & 2 & 3 & 27 \\
P3  & & 2 & 4 & 4 & 3 & 2 & 2 & 3 & 2 & 3 & 2 & 27 \\
P4  & & 4 & 4 & 4 & 3 & 2 & 3 & 3 & 4 & 3 & 4 & 34 \\
P5  & & 3 & 3 & 4 & / & 2 & 2 & 2 & 3 & 3 & 2 & 24 \\
P6  & & 2 & 3 & 4 & 2 & 1 & 4 & 2 & 2 & 3 & 4 & 27 \\
P7  & & 3 & 2 & 3 & 1 & 2 & 4 & 2 & 2 & 2 & 2 & 23 \\
P8  & & 2 & 2 & 3 & 1 & 2 & 1 & 1 & 2 & 2 & 1 & 17 \\
P9  & & 2 & 2 & 4 & 1 & 3 & 2 & 1 & 2 & 4 & 3 & 24 \\
P10 & & 3 & 2 & / & 2 & 3 & 2 & 1 & 2 & 1 & 2 & 18 \\
P11 & & 2 & 4 & 4 & 2 & 2 & 2 & 2 & 3 & 1 & 3 & 25 \\

\midrule

\multicolumn{12}{l}{\textit{Phase 2}} \\

P1  & & 2 & 2 & 3 & 1 & 2 & 3 & 2 & 3 & 4 & 4 & 26 \\
P2  & & 4 & 3 & 4 & 3 & 4 & 3 & 3 & 2 & 1 & 4 & 31 \\
P3  & & 2 & 3 & 3 & 2 & 2 & 3 & 2 & 2 & 2 & 2 & 23 \\
P4  & & 4 & 2 & 3 & 3 & 2 & 2 & 2 & 2 & 2 & 2 & 24 \\
P5  & & 2 & 2 & 2 & 1 & 1 & 1 & 2 & 2 & 2 & 2 & 17 \\
P6  & & 1 & 2 & 2 & 1 & 1 & 1 & 2 & 1 & 1 & 1 & 13 \\
P7  & & 2 & 2 & 2 & 2 & 1 & 3 & 2 & 3 & 1 & 1 & 19 \\
P8  & & 1 & 2 & 3 & 1 & 1 & 2 & 1 & 1 & 1 & 0 & 13 \\
P9  & & 4 & 0 & 4 & 3 & 3 & 3 & 3 & 3 & 4 & 2 & 29 \\
P10 & & 2 & 2 & 3 & 2 & 3 & 2 & 2 & 3 & 2 & 2 & 23 \\
P11 & & 2 & 3 & 3 & 2 & 2 & 3 & 2 & 3 & 0 & 2 & 22 \\

\bottomrule
\end{tabular}

\vspace{2mm}

\begin{minipage}{0.90\textwidth}[h!]
\footnotesize
\textit{Note.}
``/'' indicates that the participant omitted the item.
Higher total scores indicate higher perceived stress.
\end{minipage}

\end{table}

\end{document}